# Coherent Interlayer Tunneling and Negative Differential Resistance with High Current Density in Double Bilayer Graphene – WSe$_2$ Heterostructures

G. William Burg,[†] Nitin Prasad,[†] Babak Fallahazad,[†] Amithraj Valsaraj,[†] Kyounghwan Kim,[†] Takashi Taniguchi,[‡] Kenji Watanabe,[‡] Qingxiao Wang,[§] Moon J. Kim,[§] Leonard F. Register,[†] and Emanuel Tutuc*[†]

[†]Microelectronics Research Center, Department of Electrical and Computer Engineering, The University of Texas at Austin, Austin, Texas 78758, United States

[‡]National Institute for Materials Science, 1-1 Namiki Tsukuba, Ibakari 305-0044, Japan

[§]Department of Materials Science and Engineering, The University of Texas at Dallas, Richardson, Texas 75080, United States

ABSTRACT: We demonstrate gate-tunable resonant tunneling and negative differential resistance between two rotationally aligned bilayer graphene sheets separated by bilayer WSe$_2$. We observe large interlayer current densities of 2 µA/µm$^2$ and 2.5 µA/µm$^2$, and peak-to-valley ratios approaching 4 and 6 at room temperature and 1.5 K, respectively, values that are comparable to epitaxially grown resonant tunneling heterostructures. An excellent agreement between theoretical calculations using a Lorentzian spectral function for the two-dimensional (2D) quasiparticle states, and the experimental data indicates that the interlayer current stems primarily from energy and in-plane momentum conserving 2D-2D tunneling, with minimal contributions from inelastic or non-momentum-conserving tunneling. We demonstrate narrow tunneling resonances with intrinsic half-widths of 4 and 6 meV at 1.5 K and 300 K, respectively.

KEYWORDS: Coherent tunneling, negative differential resistance, graphene, $WSe_2$, heterostructure

The recent emergence of two-dimensional (2D) materials, such as graphene, hexagonal boron-nitride (hBN) and transition metal dichalcogenides (TMDs), coupled with advancing fabrication techniques for stacking 2D materials, has opened numerous pathways to explore the electronic and photonic properties, and device applications of van der Waals (vdW) heterostructures.[1] Evolving techniques for the layer-by-layer transfer of 2D materials allow for great flexibility in device structure, and have led to the study of many interesting phenomena in van der Waals heterostructures, such as quantum Hall effect[2] and moiré bands[3–6] in high mobility graphene on hBN substrates, quantum Hall effect in TMDs encapsulated in hBN[7,8], and resonant tunneling in double monolayer or double bilayer graphene separated by hBN.[9–14] The latter phenomenon requires the conservation of both electron energy and momentum in tunneling between two independently contacted 2D layers, and leads to interlayer current-voltage characteristics with gate-tunable negative differential resistance (NDR).[15]

One of the challenges in realizing functional vdW heterostructures using layer-by-layer transfers is the control of atomic registration between adjacent layers, and in particular that of rotational alignment, which is desirable for an efficient coupling between layers. In contrast to epitaxially grown heterostructures, where rotational alignment is ensured by the atomic bonding of successive layers, in vdW heterostructures of 2D materials the relative rotational alignment of different layers is most often not controlled. Because resonant tunneling requires a precise overlap of states in momentum space, and desirably a strong interlayer coupling, it serves as a powerful tool to probe the quantum fingerprints of vertical transport in vdW heterostructures. Furthermore, the gate-tunable NDR of the interlayer current-voltage characteristics enable the implementation of novel interlayer tunneling field-effect transistors (ITFETs), with applications for both Boolean and non-Boolean logic.[16–18]

In this study, we demonstrate gate-tunable resonant tunneling with a large interlayer conductance and negative differential resistance between two highly rotationally aligned bilayer graphene flakes separated by bilayer $WSe_2$. We employ four-point measurements to probe the intrinsic tunneling current-voltage characteristics independent of the contact resistance, which becomes relevant in our samples due to the large interlayer conductance. We observe current densities of 2 µA/µm$^2$ at room temperature, and 2.5 µA/µm$^2$ at 1.5 K, as well as NDR with peak-to-valley ratios (PVRs) up to 4 and 6 at room temperature and 1.5 K, respectively, which are comparable to values measured in epitaxially grown resonant tunneling heterostructures. Calculations of the tunneling current as a function of interlayer and gate bias using a simple, perturbative Hamiltonian model with Lorentzian broadening of the 2D quasiparticle states are in very good agreement with the measured tunneling current at all biasing conditions. This agreement shows the tunneling is energy and momentum conserving, and therefore coherent with respect to the single particle states.

Previous studies of resonant tunneling between two 2D layers in vdW heterostructures have used either double monolayer graphene or double bilayer graphene separated by hBN.[9–14] Because graphene has band minima at the corners of the hexagonal Brillouin zone, the crystal axes of the two mono- or bilayer graphene layers have to be rotationally aligned to ensure momentum conservation. This was done by either using the graphene flakes straight edges to identify principal crystal axes, and subsequently aligning them during transfer[10–12], or by using two mono- or bilayer graphene that stem from a single crystal domain, and are therefore rotationally aligned at the outset.[13] The use of bilayer graphene[12,13], or multilayer graphene[19], leads to narrower resonance thanks to reduced impact of the quantum capacitance. Two main drawbacks of these device designs limit applications for high speed digital electronics, and implementation beyond prototyping, desirable at wafer scale. First, the use of large bandgap hBN[20] as an interlayer dielectric reduces the interlayer current density and conductance. For example, a four-monolayer thick interlayer hBN translates into a specific interlayer conductance of ~10 nS/µm$^2$ at small interlayer bias, corresponding to an RC time constant of $10^{-6}$ s for a capacitance C = 1.8 µF/cm$^2$.[13] Second, the growth of large area hBN by chemical vapor deposition has so far resulted in lower crystal quality relative to exfoliated hBN, limiting its scalability.[21,22] While graphene double layers separated by a TMD have been reported, these samples do not show resonant tunneling[23], and the coupling of the two graphene layers through the TMD cannot be assessed.

The use of WSe$_2$ as an interlayer tunnel barrier is attractive for several reasons. First, with a bulk and monolayer bandgap of approximately 1.2 eV[24] and 2 eV[25], respectively, WSe$_2$ is a smaller bandgap alternative to the 5.8 eV[20] gap in hBN, resulting in larger tunneling currents. Additionally, WSe$_2$ can be isolated down to mono- or few-layer thick single crystals of high quality[8], which is crucial for minimizing defect induced scattering of tunneling carriers. Finally, it has been shown that in heterostructures with graphene, the mid-gap of WSe$_2$ is close to the neutrality point of graphene.[26]

Figure 1a shows a schematic of the interlayer tunneling field-effect transistor (ITFET) studied here, consisting of two individually contacted bilayer graphene flakes separated by bilayer WSe$_2$. The samples are realized using a series of dry transfers,[27] and are encapsulated with hBN as top and bottom dielectrics. The bilayer graphene flakes originate from a larger area single-crystal and remain aligned to within 0.1 degrees during the transfers, which ensures a close alignment of their crystal axes in the final heterostructure.[13] We use photoluminescence (PL) spectroscopy (Fig. 1b) to confirm the WSe$_2$ thickness.[28] Multiple contacts to each layer are defined by e-beam lithography, plasma etching, and metal deposition, which enables a decoupling of the contact resistance in vertical tunneling measurements.

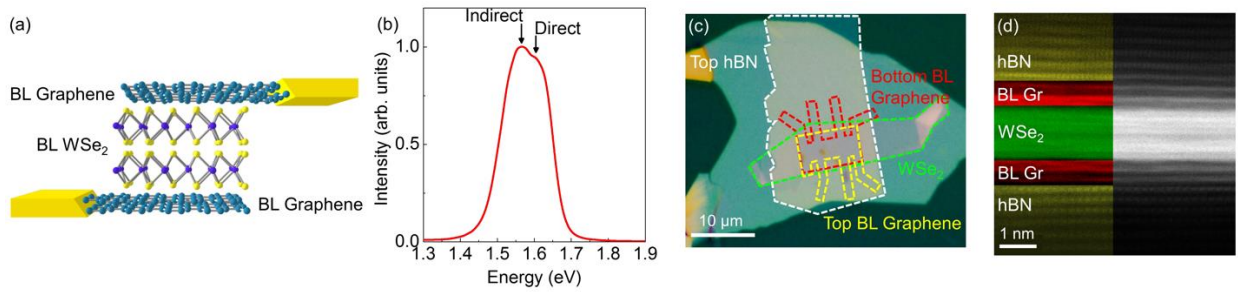

**Figure 1.** Double bilayer (BL) graphene separated by bilayer WSe$_2$ interlayer tunneling field-effect transistor. (a) Schematic representation of the device structure, including independent contacts to bilayer graphene channels. (b) PL spectrum of a WSe$_2$ bilayer used as interlayer tunnel barrier. The PL peaks corresponding to the direct and indirect gaps of bilayer WSe$_2$ are labeled. (c) Optical micrograph of a completed heterostructure. The dashed lines indicate individual layers. (d) Cross-sectional STEM of a double bilayer graphene ITFET separated by WSe$_2$, and encapsulated in hBN dielectric. The individual layers are identified using EELS and EDS.

Figure 1c shows an optical micrograph of one double bilayer graphene heterostructure separated by WSe$_2$ and encapsulated in hBN, with the contour of each layer marked. Figure 1d shows a scanning transmission electron micrograph (STEM) of the heterostructure, demonstrating atomically clean interfaces. Electron energy loss spectroscopy (EELS) and energy dispersive X-ray spectroscopy (EDS) are used to identify the different atomic layers in the heterostructure. While multiple heterostructures were fabricated for this study, we focus here on one double bilayer graphene heterostructures separated by bilayer WSe$_2$, labelled as Device #1. Data from an additional Device #2 are included in the supporting information.

We probe the tunneling current-voltage characteristics by measuring the interlayer tunneling current ($I_{int}$) as a function of the interlayer bias ($V_{TL}$) applied to the top layer, while the bottom is grounded (Figure 2b inset), at different top gate voltages ($V_{TG}$), and back gate voltage $V_{BG} = 0$ V. The top and bottom hBN dielectrics have thicknesses of 14 nm and 31 nm, respectively, corresponding to top and back gate capacitances $C_{TG} = 190$ nF/cm$^2$ and $C_{BG} = 86$ nF/cm$^2$. Figure 2a shows $I_{int}$ vs. $V_{TL}$ measured at different $V_{TG}$ values in Device #1 at room temperature. The data show clear NDR that are $V_{TG}$ dependent, with a maximum areal current density $J_{int} = 2$ µA/µm$^2$, and PVR of 3.9. Interestingly, the current peaks are followed by near discontinuous drops for all $V_{TG}$ values, a pattern that differs markedly from previously observed NDR in double layers separated by hBN, where the tunneling current has a continuous dependence on interlayer bias.[9–13] In addition, the current densities are approximately one order of magnitude larger than values corresponding to the same interlayer thickness in the best performing double layer heterostructures using hBN as the interlayer dielectric.[10] Figure 2b shows the $I_{int}$ vs. $V_{TL}$ data measured at different $V_{TG}$ in the same device at a temperature $T = 1.5$ K. Figure 2b data is very similar to Fig. 2a data, except for a slight increase in the peak current in each trace.

The biasing conditions at which resonant tunneling occurs can be understood by examining the band structures of each layer and their dependence on the applied $V_{TG}$ and $V_{TL}$. Figure 2c shows the band diagram of an ITFET for a positive $V_{TG}$ value, and at $V_{TL} = 0$ V. While the Fermi levels $\mu_{TL}$ and $\mu_{BL}$ of the top and bottom layer, respectively, are aligned, the applied gate bias induces different charge densities in each layer, leading to a finite electrostatic potential difference $V_{ES} = (\phi_{BL} - \phi_{TL})/e$ between layers, which suppresses energy and momentum conserving tunneling; here $e$ is the electron charge and $\phi_{BL}$ and $\phi_{TL}$ are the energies of the charge neutrality points (band minima) of the bottom and top layers, respectively. On the other hand, an appropriate interlayer bias restores $V_{ES} = 0$ V (Fig. 2d), and allows for energy and momentum conserving tunneling, leading to a maximum in the interlayer current. Experimentally, this can be observed by setting the $V_{TG}$ value and sweeping $V_{TL}$ in order to find the resonant condition, marked by a peak in the interlayer current.

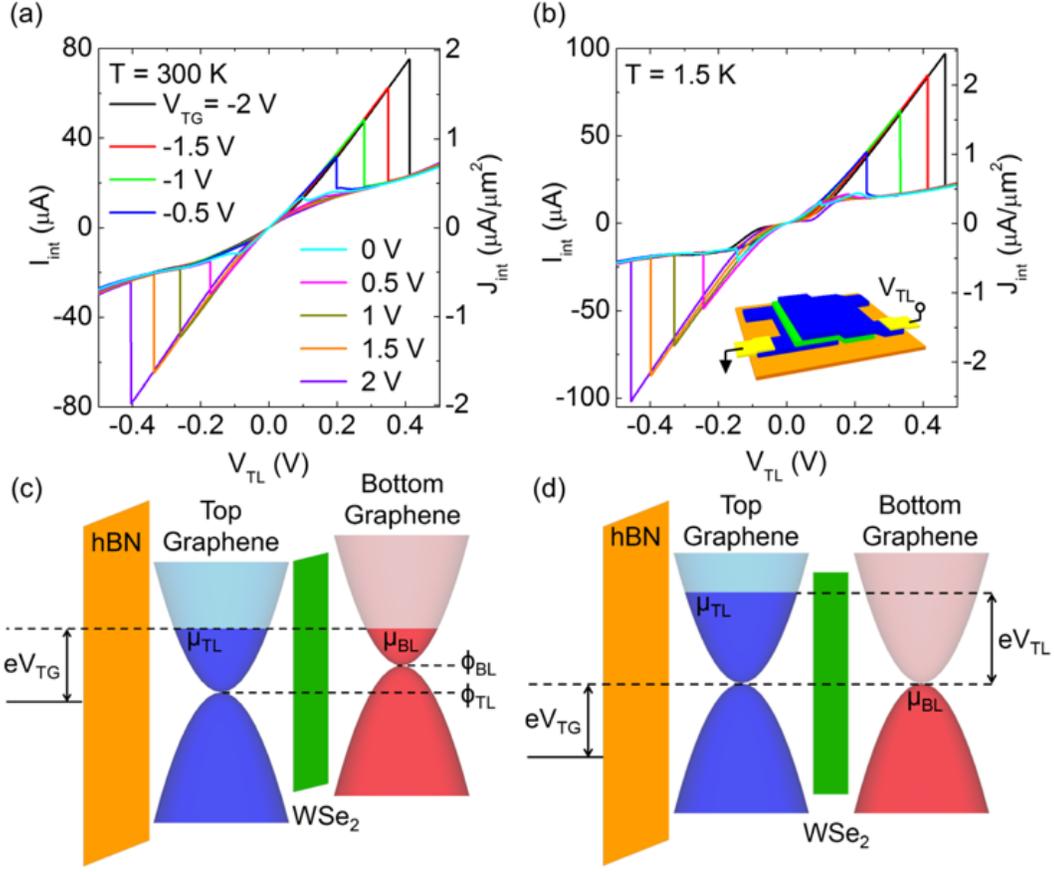

**Figure 2.** Resonant tunneling in double bilayer graphene ITFET. (a, b) Two-point $I_{int}$ vs. $V_{TL}$ at different $V_{TG}$, measured at (a) $T = 300$ K, and (b) $T = 1.5$ K in Device #1. The right axes of (a) and (b) show $I_{int}$ normalized to the bilayer graphene overlap area. Panel (b) inset shows a schematic of the interlayer biasing setup. (c, d) Simplified energy band diagrams of the ITFET at (c) $V_{TG} > 0$ V and $V_{TL} = 0$ V, and (d) aligned charge neutrality points at the same $V_{TG}$ and an appropriate finite $V_{TL}$.

To understand the discontinuity in the $I_{int}$ vs. $V_{TL}$, we consider the role of an external (contact) resistance in series with the interlayer tunneling resistance. The external resistance has contributions from both the metal/graphene contact resistance, and the in-plane resistance of the bilayer graphene extensions outside the overlap area. These external contributions are important in our devices, because Fig. 2a shows a maximum current of 80 µA at $V_{TL}$ = 0.4 V, corresponding to a total device resistance of 5 kΩ, a value comparable to that of single layer graphene field-effect transistors with similar dimensions.[27] As such, a non-negligible fraction of $V_{TL}$ drops across the contacts and bilayer graphene extensions, reducing the voltage across the tunnel barrier. This results in a "stretching" of the measured tunneling characteristic towards higher voltages. Furthermore, if the external resistance is larger than the differential resistance in the NDR region, then multiple operating points arise in the NDR region leading to discontinuities and hysteresis, as observed in Fig. 2 (a,b), and a section of the NDR current-voltage characteristic cannot be probed experimentally (see Supporting Information).

In light of these findings, a key question is what are the *intrinsic* ITFET tunneling characteristics, and how to experimentally separate these characteristics from contact resistance effects. To address this question, we perform four-point tunneling measurements where an additional pair of contacts measure the voltage across the tunnel barrier (Δ*V*), which excludes the voltage dropped across the contacts (Figure 3b inset). Figure 3a shows $I_{int}$ vs. Δ*V* measured at various $V_{TG}$ values and at room temperature in Device #1. Compared to Fig. 2 data, the resonance peaks are much sharper and appear at lower voltages relative to the two point measurements. Furthermore, consistent with the prior discussion, the sections of $I_{int}$ vs. Δ*V* data that show NDR are experimentally inaccessible as a result of finite $R_c$.

Figure 3b shows $I_{int}$ vs. *ΔV* measured at different $V_{TG}$ values, and at *T* = 1.5 K. Compared to the room temperature data of Fig. 3a, the peak (background) $I_{int}$ increase (decrease) only slightly while the peak positions are unchanged, suggesting that neither phonon scattering nor thermionic emission play a dominant role in the tunneling. Figure 3c illustrates the weak temperature dependence of $I_{int}$ vs. *ΔV* data at $V_{TG}$ = 0 V. Figure 3d shows the intrinsic differential conductance ($dI_{int}/d\Delta V$) vs. $\Delta V$ at different $T$, calculated from the Figure 3c data. The data shows narrow conductance peaks associated with the resonant tunneling, with full width at half maximums ranging from ~8 mV at $T$ = 1.5 K to 20 mV at $T$ = 300 K. The inset of Figure 3d displays a close-up of the conductance peaks of the Fig. 3c main panel. The apparent splitting of the conductance peak at $T$ = 1.5 K is likely associated with a small band gap opening in one of the two bilayers.[29] The sharp peaks in both the four-point interlayer current and differential conductance indicate a high degree of rotational alignment between layers, and suggest a high quality heterostructure with contaminant-free interfaces.

It is informative to compare the device characteristics of the double bilayer graphene separated by WSe$_2$ heterostructure to other resonant tunneling vdW and epitaxial heterostructures. We consider $J_{int}$, $\Delta V$, the specific conductance at the resonance peak, the PVR, operating temperature, and whether or not the NDR is gate-tunable as the main metrics characterizing resonant tunneling devices. Our device characteristics are comparable to many epitaxially grown heterostructures.[30–40] While some epitaxial heterostructures show a larger PVR, they typically have a lower peak specific conductance[31,34,35,38,40], and gate-tunable NDR was demonstrated only in GaAs/AlGaAs double quantum wells at temperatures lower than 170 K[35,36] (see Supporting Information). In addition, the heterostructure described here outperforms previous vdW heterostructures employing an hBN interlayer dielectric.[9–11,13]

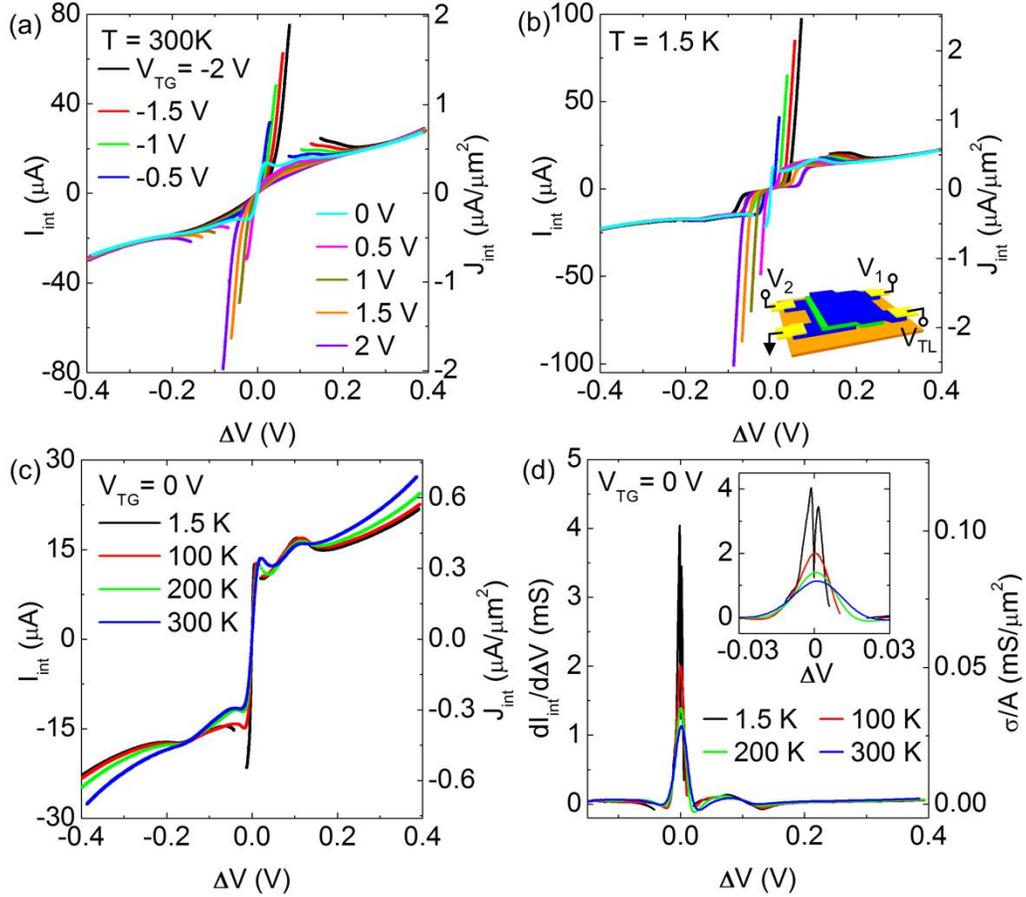

**Figure 3.** Intrinsic tunneling current-voltage characteristics. (a, b) Four-point $I_{int}$ vs. $\Delta V$ at different $V_{TG}$, measured at (a) $T = 300$ K, and (b) $T = 1.5$ K in Device #1. Panel (b) inset shows a schematic of the interlayer biasing setup. (c) Four-point $I_{int}$ vs. $\Delta V$ at $V_{TG} = 0$ V and at different $T$ values. The right axes of (a - c) show $I_{int}$ normalized to the bilayer graphene overlap area. (d) $dI_{int}/d\Delta V$ vs. $\Delta V$ corresponding to panel (c) data. The right axis shows the differential conductance ($\sigma$) normalized to the overlap area ($A$). Inset: close-up of the conductance peaks near $\Delta V = 0$.

To better understand the physics involved in the experimental tunneling characteristics, we model the ITFET using a perturbative tunneling Hamiltonian.[41,42] The band structures of the top [$\epsilon_{TL}(k)$] and bottom [$\epsilon_{BL}(k)$] bilayers are computed using a simplified tight-binding model to the leading order in wave-vector $k$ around the K point.[43] The band openings in the top and bottom bilayers are self-consistently estimated by computing the local electric fields in the bilayers, after taking electron screening into consideration.[44]

The electrostatic potential and band alignment of each graphene bilayer is computed using the following set of charge-balance equations

$$C_{IL}\left(-\frac{\phi_{TL}}{e}+\frac{\phi_{BL}}{e}\right) - C_{TG}\left(V_{TG}+\frac{\phi_{TL}}{e}\right) = Q_{TL}(\epsilon_{TL}, \mu_{TL}, \phi_{TL}) \quad (1)$$

$$C_{IL}\left(\frac{\phi_{TL}}{e}-\frac{\phi_{BL}}{e}\right) - C_{BG}\left(V_{BG}+\frac{\phi_{BL}}{e}\right) = Q_{BL}(\epsilon_{BL}, \mu_{BL}, \phi_{BL}) \quad (2)$$

where $C_{IL}$ is the interlayer capacitances per unit area, and $Q_{TL}$ ($Q_{BL}$) is the top (bottom) layer charge density.

The single particle tunneling current between the two bilayer graphene is given by

$$I_{int} = -e\int_{-\infty}^{\infty} T(E)\bigl(f(E-\mu_{TL})-f(E-\mu_{BL})\bigr)dE \quad (3)$$

where $f(E)$ is the Fermi distribution function. $T(E)$ is the vertical transmission rate of an electron at energy $E$ [45,46]:

$$T(E) = \frac{2\pi}{\hbar}\sum_{k;ss'}|t|^2 A_{TL,s}(k,E) A_{BL,s'}(k,E) \quad (4)$$

The interlayer coupling $t$ is assumed to be independent of $E$ and $k$ of the graphene bilayers. The summation is performed over all momentum states $k$ and the first two conduction and valence sub-bands, denoted by $s$ and $s'$. $A_{TL,s}$ and $A_{BL,s}$ are the spectral density functions of the band $s$ in the top and the bottom bilayers, respectively. The spectral densities are taken to be Lorentzian in form, i.e.,

$$A_s(k,E) = \frac{1}{\pi}\frac{\Gamma}{(E-\epsilon_s(k))^2+\Gamma^2}, \quad (5)$$

where Γ represents the energy broadening of the quasi-particle states, and $\epsilon_s(k)$ is the energy dispersion of band $s$ at wave-vector $k$. We note that Γ may also contain contributions from the spatial variation in the electrostatic potential difference between layers due to disorder.

The only free parameters in this model are the interlayer coupling $t$ and energy broadening parameter Γ, and the bilayers are assumed to be rotationally aligned. A rotation between the bilayers is expected to increase the broadening for small angles, and then entirely eliminate resonant tunneling at larger angles.[15] Figures 4a and 4b compare $I_{int}$ vs. $\Delta V$ calculated according to our model, to the experimental data of Figs. 3a and 3b, measured at $T$ = 300 K, and $T$ = 1.5 K, respectively. To best fit the experimental data of Figs. 3a and 3b, we use an energy broadening Γ = 6 meV at $T$ = 300 K, and Γ = 4 meV at $T$ = 1.5 K, and an interlayer coupling $|t|$ = 30 μeV.

To provide additional understanding, we perform *ab initio* density functional theory (DFT) simulations for the bilayer graphene – bilayer $WSe_2$ – bilayer graphene system. The supercell structures are relaxed using the projector-augmented wave method with a plane wave basis set as executed in the Vienna *ab initio* simulation package (VASP).[47,48] The square of the interlayer coupling is proportional to the interlayer tunneling current within a first-order approximation. The effective interlayer coupling can be estimated from DFT simulations as half of the resonant splitting in the conduction bands at zero electrostatic potential difference between the layers, where the conduction bands of the two layers would be degenerate in the absence of interlayer coupling.[49] Figure 4c shows the band structures of a bilayer graphene – bilayer $WSe_2$ – bilayer graphene heterostructure (solid, black), along with that of a single bilayer graphene (dashed, red) as reference. The relatively large energy splitting between the conduction and valence bands in the vicinity of the K point (inset) is the result of bilayer graphene coupling *to* the $WSe_2$. The smaller splitting of the conduction and valence bands within the graphene – bilayer $WSe_2$ – bilayer graphene heterostructures is the result of primary interest, and stems from coupling of the two bilayer graphene to each other *through* the bilayer of $WSe_2$. This momentum-dependent splitting is larger than 500 µeV near the band edge, and away from the K point remains substantially larger than $2|t| = 60$ µeV, used in Fig. 4(a,b) calculations. Prior theoretical work suggests that rotational misalignment between the graphene bilayers, as well as misalignment of the conductive layers with the interlayer barrier, will substantially reduce both energies.[49] The difference between the coupling determined from the experimental data and the DFT calculations suggests the interlayer current magnitude may be further improved.

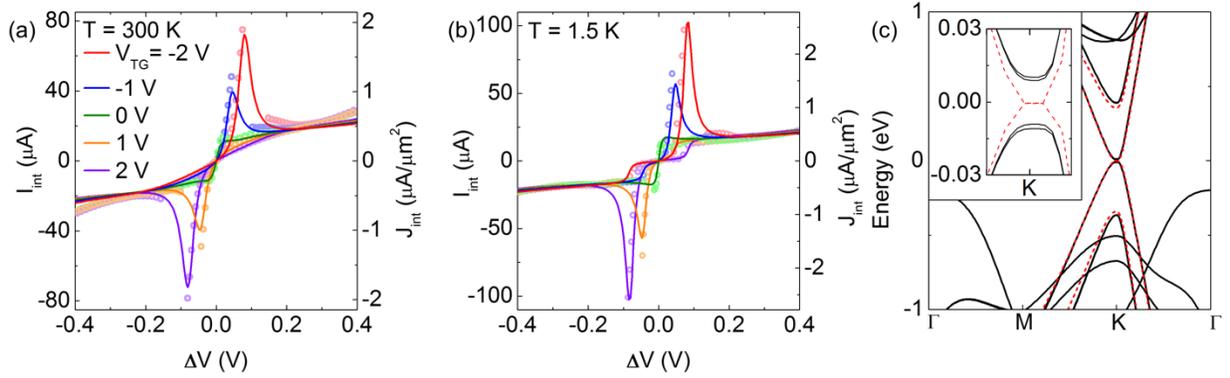

**Figure 4.** Comparison of experimental data with calculations. (a, b) Calculated $I_{int}$ vs. $\Delta V$ (solid lines) at different $V_{TG}$, at (a) $T = 300$ K and (b) $T = 1.5$ K. The experimental four-point data are shown as symbols. The model accurately reproduces the experimental findings at both temperatures with an energy broadening half width $\Gamma = 6$ meV at $T = 300$ K, and $\Gamma = 4$ meV at $T = 1.5$ K. (c) Band structures for a bilayer graphene – bilayer $WSe_2$ – bilayer graphene heterostructure (black solid lines), and bilayer graphene (dashed red lines) obtained from DFT simulations. The relatively large splitting between the conduction and valence bands in the vicinity of the K point (inset) stems from the coupling of bilayer graphene to $WSe_2$. The smaller splitting of the conduction and valence bands within the graphene – bilayer $WSe_2$ – bilayer

graphene heterostructure stems from coupling of the two bilayers of graphene through the bilayer of $WSe_2$.

It is instructive to examine in further detail key features in the $I_{int}$ vs. $\Delta V$ data, and the physical mechanisms explaining these observations. As modeled in Fig. 5a, at $T = 1.5$ K, like-band energy and momentum conserving resonant tunneling (i.e. valence to valence band, or conduction to conduction band) accounts for the resonance peak in the $I_{int}$ vs. $\Delta V$, while energy and momentum conserving non-resonant, unlike-band tunneling (e.g., conduction to valence band) produces a background tunneling current when the energy-momentum ring of intersection between unlike bands falls between the layer chemical potentials. Figure 5b shows different band alignments schematically, and the conditions leading to these different tunneling regimes, where each panel (1-6) refers to the labeled point in Fig. 5a and a corresponding voltage $V_{1-6}$. At $\Delta V = V_1$, the current is dominated by unlike-band tunneling. In this regime of operation, the current depends on the joint density of states at the ring of intersection and not on the magnitude of $\Delta V$. As we increase $\Delta V$ toward $V_2$, the ring of intersection moves outside the chemical potential difference and thereby causes a dip in the current. For the same reason, there are no states that contribute to the energy and momentum conserving current as the voltage is increased to $V_3$ and $V_4$. At $\Delta V = V_5$, the band structures of the top and bottom layers align, resulting in a large resonant current, which predominantly comes from like-band tunneling between the top and bottom layer valence bands in this case. As such, the current at resonance increases with $\Delta V$. Finally, as the voltage is further increased to $\Delta V = V_6$, energy and momentum conserving current from the top layer conduction band to bottom layer valence band takes over as the dominant source of tunneling current, with the current again dependent only on the joint density of states at the ring of intersection.

The clearly defined regions of unlike-band tunneling are less prominent as temperature increases due to the spread of the Fermi distribution, which allows tunneling between the top and bottom bilayers even when the momentum and energy conserving ring of intersection lies outside the chemical potential difference window. Furthermore, at 300 K, an additional current component appears, as shown by the small difference between the theoretical and experimental curves at large interlayer bias in Fig. 4a, suggesting a phonon-induced momentum-randomizing interlayer current between like or unlike bands. The addition of phonon scattering is qualitatively consistent with the increased broadening at 300 K. However, the predominant source of interlayer current remains energy and momentum conserving tunneling, either resonant or non-resonant. Given the excellent agreement between the modeled and experimental results, the Fig. 5 data suggest that energy and momentum conserving coherent tunneling is the dominant source of current at all interlayer biases in the experimental data at 1.5 K, a strong indication of layer-to-layer alignment, and high quality interfaces within the heterostructure.

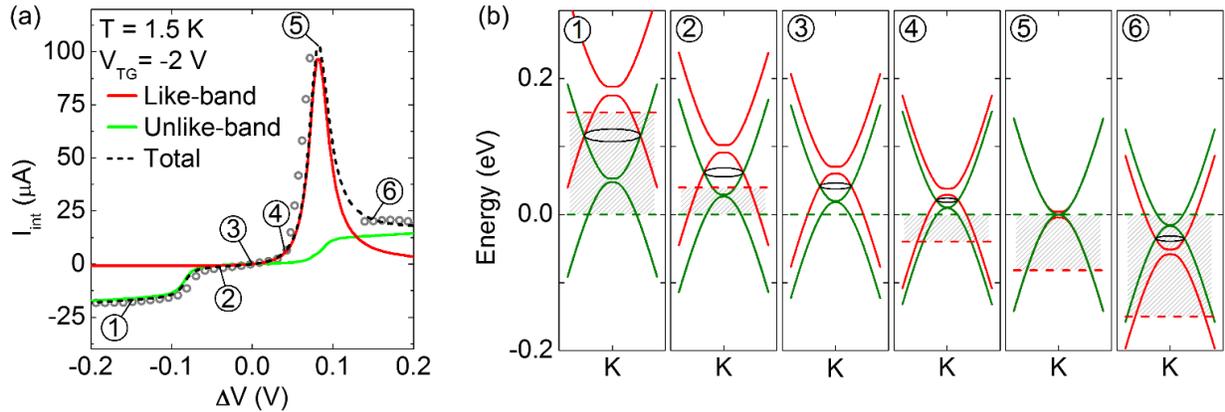

**Figure 5.** Different contributions to the total interlayer tunneling current. (a) Calculated $I_{int}$ vs. $\Delta V$ at $V_{TG}$ = -2 V and $T$ = 1.5 K. The simulated data shows the total interlayer tunneling current (black), along with the like- (red) and unlike-band (green) tunneling. The corresponding experimental data (symbols) are included for comparison. (b) Energy band-alignment of the top

(red) and bottom (green) graphene bilayers at various bias voltages. The dashed red (green) line marks the chemical potential of the top (bottom) graphene bilayer. The tunneling current between the unlike bands flows through the tunnel barrier when the intersection ring of momentum conserving states (black) lies within the chemical potential difference. Resonant tunneling current between like bands flows when the band structures completely align at $\Delta V = V_5$.

In summary, we demonstrate experimentally and model theoretically gate-tunable resonant tunneling and negative differential resistance in bilayer graphene – bilayer $WSe_2$ – bilayer graphene heterostructures. The interlayer current-voltage characteristics show current densities reaching 2 $\mu A/\mu m^2$ and 2.5 $\mu A/\mu m^2$, and PVRs of 4 and 6, at $T$ = 300 K and 1.5 K, respectively. These values coupled with narrow resonant conductance peaks suggest that heterostructures realized using layer-by-layer transfers can be of comparable quality to that of epitaxial heterostructures. The excellent agreement between theoretical calculations and experimental data indicates that the interlayer current stems primarily from energy and momentum conserving, coherent 2D-2D tunneling. We observe narrow tunneling resonances, with intrinsic half-widths of 4 and 6 meV at 1.5 K and 300 K, respectively.

ASSOCIATED CONTENT

**Supporting Information**

Discussion of contact resistance effects, and comparison of the presented device metrics with other vdW and epitaxial resonant tunneling heterostructures. This material is available free of charge at http://pubs.acs.org.

AUTHOR INFORMATION

**Corresponding Author**

*Email: etutuc@mer.utexas.edu

**Author Contributions**

G.W.B. fabricated the devices, and performed the experimental measurements, with assistance from B.F., K.K., and E.T. G.W.B., L.R.F., and E.T. analyzed the results. N.P. and A.V. performed the theoretical modeling, and analysis. Q.W. and M.J.K. provided transmission electron microscopy analysis. T.T. and K.W. supplied the hBN crystals. G.W.B., N.P., B.F., A.V., L.F.R. and E.T. contributed to the manuscript.

**Notes**

The authors declare no competing financial interest.

ACKNOWLEDGEMENTS

This work was supported by the Semiconductor Research Corp. Nanoelectronics Research Initiative SWAN center, National Science Foundation grant EECS-1610008, and Samsung Corp.

REFERENCES

(1)     Geim, A. K.; Grigorieva, I. V. *Nature* **2013**, *499* (7459), 419–425.

(2)     Dean, C. R.; Young, A. F.; Meric, I.; Lee, C.; Wang, L.; Sorgenfrei, S.; Watanabe, K.; Taniguchi, T.; Kim, P.; Shepard, K. L.; Hone, J. *Nat. Nanotechnol.* **2010**, *5* (10), 722–726.

(3)     Ponomarenko, L. A.; Gorbachev, R. V.; Yu, G. L.; Elias, D. C.; Jalil, R.; Patel, A. A.; Mishchenko, A.; Mayorov, A. S.; Woods, C. R.; Wallbank, J. R.; Mucha-Kruczynski, M.; Piot,

B. A.; Potemski, M.; Grigorieva, I. V.; Novoselov, K. S.; Guinea, F.; Fal'ko, V. I.; Geim, A. K. *Nature* **2013**, *497* (7451), 594–597.

(4)     Dean, C. R.; Wang, L.; Maher, P.; Forsythe, C.; Ghahari, F.; Gao, Y.; Katoch, J.; Ishigami, M.; Moon, P.; Koshino, M.; Taniguchi, T.; Watanabe, K.; Shepard, K. L.; Hone, J.; Kim, P. *Nature* **2013**, *497* (7451), 598–602.

(5)     Hunt, B.; Sanchez-Yamagishi, J. D.; Young, A. F.; Yankowitz, M.; LeRoy, B. J.; Watanabe, K.; Taniguchi, T.; Moon, P.; Koshino, M.; Jarillo-Herrero, P.; Ashoori, R. C. *Science* **2013**, *340* (6139), 1427–1430.

(6)     Woods, C. R.; Britnell, L.; Eckmann, A.; Ma, R. S.; Lu, J. C.; Guo, H. M.; Lin, X.; Yu, G. L.; Cao, Y.; Gorbachev, R. V.; Kretinin, A. V.; Park, J.; Ponomarenko, L. A.; Katsnelson, M. I.; Gornostyrev, Y. N.; Watanabe, K.; Taniguchi, T.; Casiraghi, C.; Gao, H.-J.; Geim, A. K.; Novoselov, K. S. *Nat. Phys.* **2014**, *10* (6), 451–456.

(7)     Cui, X.; Lee, G.-H.; Kim, Y. D.; Arefe, G.; Huang, P. Y.; Lee, C.-H.; Chenet, D. A.; Zhang, X.; Wang, L.; Ye, F.; Pizzocchero, F.; Jessen, B. S.; Watanabe, K.; Taniguchi, T.; Muller, D. A.; Low, T.; Kim, P.; Hone, J. *Nat. Nanotechnol.* **2015**, *10* (6), 534–540.

(8)     Fallahazad, B.; Movva, H. C. P.; Kim, K.; Larentis, S.; Taniguchi, T.; Watanabe, K.; Banerjee, S. K.; Tutuc, E. *Phys. Rev. Lett.* **2016**, *116* (8), 86601.

(9)     Britnell, L.; Gorbachev, R. V.; Geim, A. K.; Ponomarenko, L. A.; Mishchenko, A.; Greenaway, M. T.; Fromhold, T. M.; Novoselov, K. S.; Eaves, L. *Nat. Commun.* **2013**, *4*, 1794.

(10)    Mishchenko, A.; Tu, J. S.; Cao, Y.; Gorbachev, R. V.; Wallbank, J. R.; Greenaway, M. T.; Morozov, V. E.; Morozov, S. V.; Zhu, M. J.; Wong, S. L.; Withers, F.; Woods, C. R.; Kim,

Y.-J.; Watanabe, K.; Taniguchi, T.; Vdovin, E. E.; Makarovsky, O.; Fromhold, T. M.; Fal'ko, V. I.; Geim, A. K.; Eaves, L.; Novoselov, K. S. *Nat. Nanotechnol.* **2014**, *9* (10), 808–813.

(11)     Fallahazad, B.; Lee, K.; Kang, S.; Xue, J.; Larentis, S.; Corbet, C.; Kim, K.; Movva, H. C. P.; Taniguchi, T.; Watanabe, K.; Register, L. F.; Banerjee, S. K.; Tutuc, E. *Nano Lett.* **2015**, *15* (1), 428–433.

(12)     Kang, S.; Fallahazad, B.; Lee, K.; Movva, H.; Kim, K.; Corbet, C. M.; Taniguchi, T.; Watanabe, K.; Colombo, L.; Register, L. F.; Tutuc, E.; Banerjee, S. K. *IEEE Electron Device Lett.* **2015**, *36* (4), 405–407.

(13)     Kim, K.; Yankowitz, M.; Fallahazad, B.; Kang, S.; Movva, H. C. P.; Huang, S.; Larentis, S.; Corbet, C. M.; Taniguchi, T.; Watanabe, K.; Banerjee, S. K.; LeRoy, B. J.; Tutuc, E. *Nano Lett.* **2016**, *16* (3), 1989–1995.

(14)     Wallbank, J. R.; Ghazaryan, D.; Misra, A.; Cao, Y.; Tu, J. S.; Piot, B. A.; Potemski, M.; Pezzini, S.; Wiedmann, S.; Zeitler, U.; Lane, T. L. M.; Morozov, S. V.; Greenaway, M. T.; Eaves, L.; Geim, A. K.; Fal'ko, V. I.; Novoselov, K. S.; Mishchenko, A. *Science* **2016**, *353* (6299), 575–579.

(15)     Feenstra, R. M.; Jena, D.; Gu, G. *J. Appl. Phys.* **2012**, *111* (4), 43711.

(16)     Zhao, P.; Feenstra, R. M.; Gu, G.; Jena, D. *IEEE Trans. Electron Devices* **2013**, *60* (3), 951–957.

(17)     Banerjee, S. K.; Register, L. F.; Tutuc, E.; Reddy, D.; MacDonald, A. H. *IEEE Electron Device Lett.* **2009**, *30* (2), 158–160.

(18)    Sedighi, B.; Hu, X. S.; Nahas, J. J.; Niemier, M. *IEEE J. Emerg. Sel. Top. Circuits Syst.* **2014**, *4* (4), 438–449.

(19)    Kang, S.; Prasad, N.; Movva, H. C. P.; Rai, A.; Kim, K.; Mou, X.; Taniguchi, T.; Watanabe, K.; Register, L. F.; Tutuc, E.; Banerjee, S. K. *Nano Lett.* **2016**, *16* (8), 4975–4981.

(20)    Watanabe, K.; Taniguchi, T.; Kanda, H. *Nat. Mater.* **2004**, *3* (6), 404–409.

(21)    Shi, Y.; Hamsen, C.; Jia, X.; Kim, K. K.; Reina, A.; Hofmann, M.; Hsu, A. L.; Zhang, K.; Li, H.; Juang, Z.-Y.; Dresselhaus, M. S.; Li, L.-J.; Kong, J. *Nano Lett.* **2010**, *10* (10), 4134–4139.

(22)    Kim, K. K.; Hsu, A.; Jia, X.; Kim, S. M.; Shi, Y.; Hofmann, M.; Nezich, D.; Rodriguez-Nieva, J. F.; Dresselhaus, M.; Palacios, T.; Kong, J. *Nano Lett.* **2012**, *12* (1), 161–166.

(23)    Georgiou, T.; Jalil, R.; Belle, B. D.; Britnell, L.; Gorbachev, R. V.; Morozov, S. V.; Kim, Y.-J.; Gholinia, A.; Haigh, S. J.; Makarovsky, O.; Eaves, L.; Ponomarenko, L. A.; Geim, A. K.; Novoselov, K. S.; Mishchenko, A. *Nat. Nanotechnol.* **2013**, *8* (2), 100–103.

(24)    Kam, K. K.; Parkinson, B. A. *J. Phys. Chem.* **1982**, *86* (4), 463–467.

(25)    Zhang, C.; Chen, Y.; Johnson, A.; Li, M.-Y.; Li, L.-J.; Mende, P. C.; Feenstra, R. M.; Shih, C.-K. *Nano Lett.* **2015**, *15* (10), 6494–6500.

(26)    Kim, K.; Larentis, S.; Fallahazad, B.; Lee, K.; Xue, J.; Dillen, D. C.; Corbet, C. M.; Tutuc, E. *ACS Nano* **2015**, *9* (4), 4527–4532.

(27)    Wang, L.; Meric, I.; Huang, P. Y.; Gao, Q.; Gao, Y.; Tran, H.; Taniguchi, T.; Watanabe, K.; Campos, L. M.; Muller, D. A.; Guo, J.; Kim, P.; Hone, J.; Shepard, K. L.; Dean, C. R. *Science* **2013**, *342* (6158), 614–617.

(28)    Terrones, H.; Corro, E. D.; Feng, S.; Poumirol, J. M.; Rhodes, D.; Smirnov, D.; Pradhan, N. R.; Lin, Z.; Nguyen, M. a. T.; Elías, A. L.; Mallouk, T. E.; Balicas, L.; Pimenta, M. A.; Terrones, M. *Sci. Rep.* **2014**, *4*, 4215.

(29)    Zhang, Y.; Tang, T.-T.; Girit, C.; Hao, Z.; Martin, M. C.; Zettl, A.; Crommie, M. F.; Shen, Y. R.; Wang, F. *Nature* **2009**, *459* (7248), 820–823.

(30)    Huang, C. I.; Paulus, M. J.; Bozada, C. A.; Dudley, S. C.; Evans, K. R.; Stutz, C. E.; Jones, R. L.; Cheney, M. E. *Appl. Phys. Lett.* **1987**, *51* (2), 121–123.

(31)    Söderström, J. R.; Chow, D. H.; McGill, T. C. *Appl. Phys. Lett.* **1989**, *55* (11), 1094–1096.

(32)    Smet, J. H.; Broekaert, T. P. E.; Fonstad, C. G. *J. Appl. Phys.* **1992**, *71* (5), 2475–2477.

(33)    Brown, E. R.; Söderström, J. R.; Parker, C. D.; Mahoney, L. J.; Molvar, K. M.; McGill, T. C. *Appl. Phys. Lett.* **1991**, *58* (20), 2291–2293.

(34)    Day, D. J.; Yang, R. Q.; Lu, J.; Xu, J. M. *J. Appl. Phys.* **1993**, *73* (3), 1542–1544.

(35)    Blount, M. A.; Simmons, J. A.; Moon, J. S.; Baca, W. E.; Reno, J. L.; Hafich, M. J. *Semicond. Sci. Technol.* **1998**, *13* (8A), A180.

(36)    Simmons, J. A.; Blount, M. A.; Moon, J. S.; Lyo, S. K.; Baca, W. E.; Wendt, J. R.; Reno, J. L.; Hafich, M. J. *J. Appl. Phys.* **1998**, *84* (10), 5626–5634.

(37)    Duschl, R.; Schmidt, O. G.; Eberl, K. *Appl. Phys. Lett.* **2000**, *76* (7), 879–881.

(38)    Watanabe, M.; Funayama, T.; Teraji, T.; Sakamaki, N. *Jpn. J. Appl. Phys.* **2000**, *39* (7B), L716.

(39) Suda, Y.; Koyama, H. *Appl. Phys. Lett.* **2001**, *79* (14), 2273–2275.

(40) Rommel, S. L.; Pawlik, D.; Thomas, P.; Barth, M.; Johnson, K.; Kurinec, S. K.; Seabaugh, A.; Cheng, Z.; Li, J. Z.; Park, J. S.; Hydrick, J. M.; Bai, J.; Carroll, M.; Fiorenza, J. G.; Lochtefeld, A. In Proceedings of the IEEE International Electron Devices Meeting, San Francisco, USA, Dec 15−17, 2008.

(41) Zheng, L.; MacDonald, A. H. *Phys. Rev. B* **1993**, *47* (16), 10619–10624.

(42) Turner, N.; Nicholls, J. T.; Linfield, E. H.; Brown, K. M.; Jones, G. A. C.; Ritchie, D. A. *Phys. Rev. B* **1996**, *54* (15), 10614–10624.

(43) Castro Neto, A. H.; Guinea, F.; Peres, N. M. R.; Novoselov, K. S.; Geim, A. K. *Rev. Mod. Phys.* **2009**, *81* (1), 109–162.

(44) Min, H.; Sahu, B.; Banerjee, S. K.; MacDonald, A. H. *Phys. Rev. B* **2007**, *75* (15), 155115.

(45) Bardeen, J. *Phys. Rev. Lett.* **1961**, *6* (2), 57–59.

(46) Barrera, S. C. de la; Gao, Q.; Feenstra, R. M. *J. Vac. Sci. Technol. B* **2014**, *32* (4), 04E101.

(47) Kresse, G.; Furthmüller, J. *Phys. Rev. B* **1996**, *54* (16), 11169–11186.

(48) Kresse, G.; Furthmüller, J. *Comput. Mater. Sci.* **1996**, *6* (1), 15–50.

(49) Valsaraj, A.; Register, L. F.; Tutuc, E.; Banerjee, S. K. *J. Appl. Phys.* **2016**, *120* (13), 134310.